# Piezooptic Coefficients and Acoustic Wave Velocities in $Sn_2P_2S_6$ Crystals

[1]O.Mys, [1]I.Martynyuk-Lototska, [2]A.Grabar, [2]Yu.Vysochanskii and [1]R.Vlokh

[1]Institute of Physical Optics, 23 Dragomanov St., 79005 Lviv, Ukraine, E-mail: vlokh@ifo.lviv.ua
[2] Institute for Solid State Physics and Chemistry, Uzhgorod National University, 54 Voloshyn St., 88000 Uzhgorod, Ukraine, E-mail: agrabar@univ.uzhgorod

**Abstract**
Piezooptic coefficients of $Sn_2P_2S_6$ crystals are experimentally determined for $\lambda = 632.8 nm$ and $T = 293 K$ with the aid of interferometric technique. The components of the elastic stiffness tensor for these crystals are calculated on the basis of studies for the acoustic wave velocities. It is shown that acoustooptic figure of merit can achieve extremely high values for $Sn_2P_2S_6$ crystals ($M_2 \sim 2\times10^{-12} s^3/kg^2$).

**Introduction**

$Sn_2P_2S_6$ crystals belong to semiconductor-ferroelectric materials with nontrivial properties that make them attractive for various practical applications, particularly in optoelectronics. Though electro-, piezo- and magnetooptic coefficients represent the most important parameters of any material, which are crucial in considering the very possibility of its utilization in devices for controlling optical radiation, those coefficients are still not completely studied for $Sn_2P_2S_6$ crystals. Moreover, the above compound represents a wide family of semiconductors $Sn_{2y}Pb_{2(1-y)}P_2S_{6x}Se_{6(1-x)}$ (see, e.g., [1]), whose parameters may be efficiently governed when substituting relevant chemical components. Specifically, such basic properties as the refractive indices and the bandgap energy have proven to be sensitive to this substitution. This could make the mentioned crystal family the most attractive materials for the integrated optics, especially if one takes the expected possibility for growing epitaxial layers into consideration.

$Sn_2P_2S_6$ crystals manifest a proper second-order paraelectric-to-ferroelectric phase transition at 337 K with the change of point symmetry group $2/mFm$, while $Sn_2P_2Se_6$ is characterized with the phase transition into a proper ferroelectric phase with the symmetry $m$ ($T_c$=193 K). These crystals are semiconductors, with the bandgap energy $E_g$=1.7 eV at $T$=293 K [2].

The relevant studies of optical effects in $Sn_2P_2S_6$ crystals induced by different external fields have been originated from the estimations of electrooptic (EO) coefficients on the basis of temperature variations of optical refraction caused by spontaneous electric polarization [3]. Later on, temperature dependence of the effective EO coefficient has been studied in the work [4], where high enough EO figure of merit of these crystals has been detected for the first time. Finally, a detailed study of EO effect in $Sn_2P_2S_6$ crystals has been performed in [5]. It has been shown that the EO coefficient achieves extremely high values: $r_{11}=1.74\times10^{-10}$ m/V at the room temperature and $\lambda=632.8$ nm, while at the phase transition temperature it becomes still larger ($4.5\times10^{-9}$ m/V). One can expect that such the high values of EO coefficients in $Sn_2P_2S_6$ crystals could be caused by a secondary piezooptic (PO) effect due to piezoelectric effect. Nevertheless, as far as we know, the EO data for $Sn_2P_2S_6$ [3–5] are almost the only available in the literature for $Sn_{2y}Pb_{2(1-y)}P_2S_{6x}Se_{6(1-x)}$ family, in what the effects of optical birefringence induced by external fields are concerned. At least we can only remind the results for the effective PO coefficient of $Sn_2P_2S_6$ crystals presented in the study [6], which has confirmed that the coefficient is indeed high. The PO coefficient reported in [1] is also high ($\pi_{66}=5\times10^{-11}$ m$^2$/N) and the same is also true of the corresponding acoustooptic (AO) figure of merit estimated in this work ($M_2=540\times10^{-15}$ s$^3$/kg). It is worthwhile that the PO (or elastooptic) coefficients and the ultrasonic wave velocities represent the data necessary for predicting practical applicability of material in acoustooptics.

Following from the mentioned above, we have undertaken detailed studies for the PO, elastooptic, acoustic and AO effects in the crystals of $Sn_{2y}Pb_{2(1-y)}P_2S_{6x}Se_{6(1-x)}$ family. We begin from the PO coefficients and the ultrasonic velocities for $Sn_2P_2S_6$ crystals, which are presented in this work.

**Experimental procedure and results**

Synthesis of $Sn_2P_2S_6$ compounds and the growth of single crystals have been performed on the basis of chemical vapour transport reactions [7]. In order to implement PO experiments, after X-ray orientation procedure we have prepared a sample with the shape of parallelepiped ($5.43\times5.08\times4.13$ $mm^3$) and the faces perpendicular to principal crystallographic directions for the light propagation and the application of compressive mechanical stresses. All the experimental measurements have been performed on single-domain samples. It is necessary to remind that $Sn_2P_2S_6$ crystals belong to low-symmetry monoclinic point group. Thus, the orientation of optical indicatrix (OI) depends on the wavelength, temperature, etc. Y axis of the crystallographic coordinate system coincides with the corresponding axis of the OI, while X and Z axes of the latter are rotated by $\simeq 45°$ in XZ plane with respect to crystallographic directions (at the room temperature and the wavelength of radiation $632.8 nm$). The form of the PO tensor for the point symmetry group $m$ is as follows:

$$
\begin{array}{c|cccccc}
 & \sigma_1 & \sigma_2 & \sigma_3 & \sigma_4 & \sigma_5 & \sigma_6 \\
\hline
\Delta B_1 & \pi_{11} & \pi_{12} & \pi_{13} & 0 & \pi_{15} & 0 \\
\Delta B_2 & \pi_{21} & \pi_{22} & \pi_{23} & 0 & \pi_{25} & 0 \\
\Delta B_3 & \pi_{31} & \pi_{32} & \pi_{33} & 0 & \pi_{35} & 0 \\
\Delta B_4 & 0 & 0 & 0 & \pi_{44} & 0 & \pi_{46} \\
\Delta B_5 & \pi_{51} & \pi_{52} & \pi_{53} & 0 & \pi_{55} & 0 \\
\Delta B_6 & 0 & 0 & 0 & \pi_{54} & 0 & \pi_{66}
\end{array} \quad (1)
$$

It does not depend on the orientation of OI with respect to the axes of crystallographic system, unlike the magnitude of PO coefficients. Thus, for avoiding ambiguity in the determination of PO coefficients we have used the crystallographic coordinate system, with a possibility of further recalculation of the coefficient to the principal coordinate system (i.e., the system of the OI) after all the coefficients in Eq. (1) are determined.

It is seen that the matrix (1) is rather complicated and includes twenty independent coefficients. In the present paper we represent the results for the nine of them: $\pi_{11}$, $\pi_{12}$, $\pi_{13}$, $\pi_{21}$, $\pi_{22}$, $\pi_{23}$, $\pi_{31}$, $\pi_{32}$ and $\pi_{33}$. Let us consider equation for the OI perturbed by the mechanical stress $\sigma_2$ and written in the crystallographic system:

$$(B_1 + \pi_{12}\sigma_2)X^2 + (B_2 + \pi_{22}\sigma_2)Y^2 + (B_3 + \pi_{32}\sigma_2)Z^2 + 2\pi_{52}\sigma_2 ZX = 1. \quad (2)$$

One can find that the mechanical stress leads to a small rotation of OI around the Y axis:

$$\tan 2\xi_Y = \frac{\pi_{52}\sigma_2}{B_1 - B_3 + (\pi_{12} - \pi_{32})\sigma_2}, \quad (3)$$

which is of the order of $10^{-3} - 10^{-4} rad$. This small OI rotation cannot essentially affect the accuracy of determination of $\pi_{12}$, $\pi_{22}$ and $\pi_{32}$, when the incident light is polarized parallel to X, Y and Z axis, respectively. The same is true for the other components of compressive stresses. Thus, one can neglect the influence of the components $\pi_{51}$, $\pi_{52}$ and $\pi_{53}$. However, it is necessary to remember that the mechanical stress components $\sigma_1$ and $\sigma_3$ in the crystallographic system correspond to the components $\sigma_5$ and $-\sigma_5$ in the coordinate system of OI.

We studied the PO effect with interferometric technique based on employing Mach-Zehnder interferometer (see Fig. 1) and determining optical retardation that corresponds to a half-wave mechanical stress $^{\lambda/2}\sigma_m$.

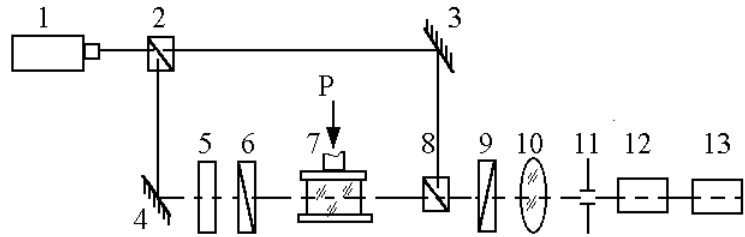

Fig. 1. Experimental setup for PO measurements: 1 – He-Ne laser, 2, 8 – semi-reflecting mirror cubes, 3, 4 – mirrors, 5 – $\lambda/4$ plate, 6, 9 – polarizers, 7 – setup for application of mechanical stresses, 10 – objective lens, 11 – diaphragm, 12 – photodetector, 13 – multimeter.

The PO coefficients were calculated with the formula (see, e.g., [8])

$$\pi_{im} = \frac{2\delta\Delta_k}{^{\lambda/2}\sigma_m n_i^3 d_k}, \qquad (4)$$

where $\delta\Delta_k$ is the optical retardation change, $n_i$ the refractive index for $i-th$ polarization of the incident light and $d_k$ the sample thickness along the direction of light propagation.

The accuracy for determination of the PO coefficients directly obtained from the experiment was equal to 15%. The measurements were performed under constant field conditions for the case of He-Ne laser (the wavelength of radiation $632.8 nm$) and the temperature $T = 393K$. The experimental geometry and the values of PO coefficients are presented in Table 1.

Table 1. Experimental geometry and PO coefficients of $Sn_2P_2S_6$ crystals.

| Light wave vector direction | Polarization of incident light, mechanical stress component | Module of PO coefficient, $pm^2/N$ |
|---|---|---|
| X | Z, $\sigma_2$ | $\pi_{32} = 9.94$ |
| X | Y, $\sigma_2$ | $\pi_{22} = 8.48$ |
| Z | X, $\sigma_2$ | $\pi_{12} = 4.63$ |
| X | Y, $\sigma_1$ | $\pi_{21} = 8.8$ |
| Y | Z, $\sigma_1$ | $\pi_{31} = 13.7$ |
| Y | X, $\sigma_1$ | $\pi_{11} = 15.6$ |
| X | Z, $\sigma_3$ | $\pi_{33} = 2.61$ |
| Y | X, $\sigma_3$ | $\pi_{13} = 2.2$ |

| X | Y, $\sigma_3$ | $\pi_{23} = 1.49$ |

The measurements of the velocity of longitudinal and transverse ultrasonic waves were performed on single crystals by the pulse-echo overlap method. The accuracy of the velocity measurements was about 0.5%. However the total error for the absolute value of acoustic wave velocities was higher and depended also on the misalignment of sample orientation. The acoustic waves in samples were excited, using LiNbO$_3$ transducers (the resonance frequency $f$ =10 MHz, the bandwidth $\Delta f$ =0.1 MHz and the acoustic power $P_a$=1–2 W). Notice that the acoustic velocities and the elastic modules are presented below for the crystallographic system but not the principal coordinate system of the elastic modules.

Some of the ultrasonic waves velocities for Sn$_2$P$_2$S$_6$ crystals and all the components of elastic stiffness tensor have previously been determined with different methods, including the pulse echo overlap method, Brillouin scattering spectroscopy and the neutron scattering [1,9–12]. Large dispersion of these data is probably caused by the use of samples grown with different techniques and a low accuracy of sample orientation. The accuracy of these measurements is not satisfactory for AO applications of the mentioned crystals. Our results for the ultrasonic wave velocities are presented in Table 2. These data agree more or less with the results published previously. It is necessary to note that, since the sign of the coordinate system is not determined in the present measurements, the wave velocities for the acoustic waves, whose polarization or propagation directions are not parallel to the principal crystallographic axes, are determined with the accuracy up to the sign of coordinate system. We have also observed the dependence of acoustic velocities on pre-illuminating conditions. The results presented below are obtained for the non-illuminated samples.

Table 2. Ultrasonic wave velocities for Sn$_2$P$_2$S$_6$ crystals at $T = 293K$.

| $V_{ij}$ | Direction of wave propagation | Wave displacement vector direction | Velocity, $m/s$ |
|---|---|---|---|
| $V_{11}$ | [100] | [100] | $3580 \pm 40$ |
| $V_{22}$ | [010] | [010] | $2870 \pm 40$ |
| $V_{33}$ | [001] | [001] | $3600 \pm 20$ |
| $V_{23}$ | [010] | [001] | $2610 \pm 40$ |
| $V_{21}$ | [010] | [100] | $2100 \pm 20$ |
| $V_{32}$ | [001] | [010] | $2500 \pm 50$ |
| $V_{31}$ | [001] | [100] | $2880 \pm 60$ |

| $V_{12}$ | [100] | [010] | $2220 \pm 50$ |
|---|---|---|---|
| $V_{13}$ | [100] | [001] | $2470 \pm 25$ |
| $V_{55}$ | [101] | [101] | $3890 \pm 20$ |
| $V_{\bar{5}\bar{5}}$ | [$\bar{1}0\bar{1}$] | [$\bar{1}0\bar{1}$] | $3870 \pm 20$ |
| $V_{5\bar{5}}$ | [101] | [$\bar{1}0\bar{1}$] | $2260 \pm 20$ |
| $V_{52}$ | [101] | [010] | $2200 \pm 20$ |
| $V_{\bar{5}5}$ | [$\bar{1}0\bar{1}$] | [101] | $2460 \pm 25$ |
| $V_{\bar{5}2}$ | [$\bar{1}0\bar{1}$] | [010] | $1740 \pm 20$ |
| $V_{44}$ | [011] | [011] | $3150 \pm 30$ |
| $V_{4\bar{4}}$ | [011] | [$0\bar{1}\bar{1}$] | $2580 \pm 30$ |
| $V_{41}$ | [011] | [100] | $1890 \pm 20$ |
| $V_{66}$ | [110] | [110] | $4060 \pm 40$ |
| $V_{\bar{6}\bar{6}}$ | [$\bar{1}\bar{1}0$] | [$\bar{1}\bar{1}0$] | $3910 \pm 40$ |
| $V_{6\bar{6}}$ | [110] | [$\bar{1}\bar{1}0$] | $2290 \pm 20$ |
| $V_{\bar{6}6}$ | [$\bar{1}\bar{1}0$] | [110] | $2270 \pm 20$ |
| $V_{63}$ | [110] | [001] | $1560 \pm 20$ |
| $V_{\bar{6}3}$ | [$\bar{1}\bar{1}0$] | [001] | $1560 \pm 20$ |

On the basis of Christoffel equation we have calculated some components of the elastic stiffness tensor: $C_{11} = 4.5 \times 10^{10} N/m^2$, $C_{22} = 2.9 \times 10^{10} N/m^2$, $C_{33} = 4.5 \times 10^{10} N/m^2$, $C_{12} \approx 4.4 \times 10^{10} N/m^2$ (with using of $V_{66}$ at calculation), $C_{13} \approx 0.8 \times 10^{10} N/m^2$ (with using of $V_{55}$ and $V_{5\bar{5}}$ at calculation), $C_{23} \approx 1.1 \times 10^{10} N/m^2$ (with using of $V_{41}$ at calculation) $C_{44} = 2.2 \times 10^{10} N/m^2$, $C_{55} = 2.2 \times 10^{10} N/m^2$, $C_{66} = 1.8 \times 10^{10} N/m^2$, $C_{46} = 0.4 \times 10^{10} N/m^2$, $C_{35} = 0.5 \times 10^{10} N/m^2$, $C_{15} = 0.2 \times 10^{10} N/m^2$.

**Conclusions**

The results presented in this paper give rise to extensive investigations of piezo-, elasto- and AO properties of $Sn_{2y}Pb_{2(1-y)}P_2S_{6x}Se_{6(1-x)}$ crystals. The results obtained by us show that the velocities of the chosen acoustic waves in $Sn_2P_2S_6$ crystals are quite low. This fact, together with the large PO coefficients (some of them are of the order of $10\, pm^2/N$) and large refractive indices, enables one to predict extremely high AO figures of merit for the above crystals. For example, if we estimate the $M_2$ value on the basis of the lowest acoustic wave velocity $V = 1560 m/s$ and the other averaged data ($n \approx 3.0$, $\pi \approx 7 pm^2/N$, $C \approx 2.8 \times 10^{10} N/m^2$ and $p \approx 0.2$), we get $M_2 \sim 2 \times 10^{-12} s^3/kg^2$. This is at least more than two times larger than the corresponding value for the best AO material, $TeO_2$ crystals. However, determination of complete matrices of the PO and elastooptic coefficients, as well as the elastic stiffness coefficients and elastic compliances, would be necessary for precise enough evaluation of the AO figure of merit and optimization of AO interaction geometry. The appropriate results will be a subject of our forthcoming papers.

**Acknowledgement**

Some of the authors (O.M., I.M.-L. and R.V.) acknowledge financial support of this study from the Ministry of Education and Science of Ukraine (the Project N0106U000616).